\documentclass{article}

\usepackage{arxiv}
\usepackage[utf8]{inputenc} 
\usepackage[T1]{fontenc}    
\usepackage{hyperref}    
\usepackage{url}          
\usepackage{booktabs}     
\usepackage{amsfonts}     
\usepackage{nicefrac}     
\usepackage{microtype} 
\usepackage{lipsum}
\usepackage{graphicx}
\usepackage{doi}
\usepackage{amsthm}
\usepackage{multirow}
\usepackage{longtable}
\usepackage{amsmath,bm}
\usepackage{siunitx}
\usepackage{subfig}
\usepackage{soul}

\newcommand{\sig}[1]{\ifthenelse{\equal{#1}{0.1}}{$^{*}$}{\ifthenelse{\equal{#1}{0.05}}{$^{**}$}{\ifthenelse{\equal{#1}{0.01}}{$^{***}$}{}}}}

\title{Persistence is All You Need -- A Topological \\ Lens on Microstructural Characterization}

\author{
    \href{https://orcid.org/0009-0005-0510-7610}{\includegraphics[scale=0.06]{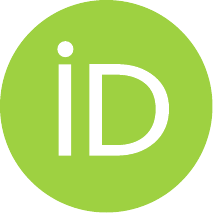}
    \hspace{1mm}Maksym Szemer} \\
    AGH University of Krakow, Krakow, Poland \\
    \texttt{szemermaksym@student.agh.edu.pl} \\
	\And
	\href{https://orcid.org/0000-0002-2959-3044}{\includegraphics[scale=0.06]{orcid.pdf}
    \hspace{1mm}Szymon Buchaniec} \\
    AGH University of Krakow, Krakow, Poland \\
    \texttt{buchaniec@agh.edu.pl} \\
    \And
	\href{https://orcid.org/0000-0003-4911-5880}{\includegraphics[scale=0.06]{orcid.pdf}
    \hspace{1mm}Grzegorz Brus} \thanks{Corresponding author: Grzegorz Brus brus@agh.edu.pl} \\
    AGH University of Krakow, Krakow, Poland \\
    \texttt{brus@agh.edu.pl} \\
}

\renewcommand{\shorttitle}{\textit{arXiv} Template}

\hypersetup{
pdftitle={Persistence is All You Need},
pdfsubject={cs.LG, cs.AI, math.AT},
pdfauthor={Maksym ~Szemer, Szymon ~Buchaniec, Grzegorz ~Brus},
pdfkeywords={Deep Learning, Topological Data Analysis, Persistent Homology, Solid Oxide Fuel Cell, Microstructure, Computational Topology},
}

\begin{document}
\maketitle

\begin{abstract}
	The microstructure critically governs the properties of materials used in energy and chemical engineering technologies, from catalysts and filters to thermal insulators and sensors. Therefore, accurate design is based on quantitative descriptors of microstructural features. Here we show that eight key descriptors can be extracted by a single workflow that fuses computational topology with assembly–learning–based regression. First, 1312 synthetic three-dimensional microstructures were generated and evaluated using established algorithms, and a labeled data set of ground-truth parameters was built. Converting every structure into a persistence image allowed us to train a deep neural network that predicts the eight descriptors. In an independent test set, the model achieved on average $R^2$ $\approx 0.84$ and Pearson $r\approx 0.92$, demonstrating both precision and generality. The approach provides a unified and scalable tool for rapid characterization of functional porous materials.
\end{abstract}

% keywords can be removed

\section{Introduction}
Numerous phenomena in science and engineering are affected by the system's features at the microscale. Porous materials used for filtering\cite{Peng2024}, polycrystalline structures in semiconductors\cite{Ren2025}, knitted spacer fabrics\cite{Li2024concrete} and many more have a crucial impact on the effectiveness, durability, and manufacturing properties of the designed technology. The microstructure of the selected material must be designed for a specific application, which is often a complex task that involves manufacturing and modeling challenges. Microstructures are often described quantitatively to simplify modeling and give insight into the dependency between system performance and microstructure\cite{Iwai.2010}. In the context of the case study analyzed in this article, solid oxide cells (SOCs) represent one of the most promising technologies for clean and efficient energy conversion and storage. The performance and durability of these devices are fundamentally linked to their complex electrode microstructures\cite{Brus2017}. Characterizing critical microstructural parameters, such as phase volume fractions, particle size, connectivity, tortuosity factors, and triple phase boundary (TPB) length density is essential for optimizing electrode functionality\cite{Buchaniec.2019,Yan2019c4}

Contemporary imaging and processing techniques allow direct observation, visualization, and quantification of the microstructure\cite{Joos2010}. Early studies relied on 2D microscopy, while modern approaches use 3D tomography to capture the complex geometry of all three phases. Focused Ion Beam–Scanning Electron Microscopy (FIB-SEM) tomography is among the most commonly used technologies. It is a destructive imaging method where a focused ion beam sequentially mills thin layers of the sample material. After each milling step, a SEM image, typically obtained using an in-lens detector, is captured. By stacking these serial images, a detailed 3D volume of the microstructure is reconstructed with voxel resolutions commonly in the tens of nanometers range \cite{Iwai.2010, Lu.2017}. FIB-SEM tomography provides sufficient contrast to clearly resolve typical SOC particles such as Ni, YSZ, GDC or LSCF \cite{Holzer.2011, Wilson.2006}. The phase segmentation process usually involves grayscale thresholding and, in some cases, energy-dispersive spectroscopy (EDS) for phase identification \cite{Wilson.2006, Lu.2017}. Quantitative microstructural analyses directly derived from these 3D datasets include: phase volume fractions, calculated by voxel counting, particle size distributions, obtainable through various methods such as PSD\textsubscript{D}, PSD\textsubscript{C} or Intercept Method. In PSD\textsubscript{D} particle sizes are quantified as equivalent diameters of individual volumes segmented from the interconnected 3D microstructure \cite{Holzer.2011}. Analogically in PSD\textsubscript{C} a morphological method that determines particle size by evaluating the volumes accessible to spheres of decreasing radius \cite{munch.2008}. Finally, in the intercept method, particle sizes are estimated by measuring intersection lengths between particles and randomly oriented test lines through the 3D structure \cite{Iwai.2010}. Connectivity is evaluated through cluster analyses to determine if individual phases form interconnected networks \cite{Grew.2010,Kishimoto.2011p9d}. Triple-phase boundary length density is often computed using a combination of such morphological algorithms as the phase-expansion and voxel-centroid methods \cite{Shikazono.2010, Iwai.2010}. The tortuosity factor is determined by simulating transport phenomena \cite{Wei.2015,Prokop2023}, Lattice Boltzmann method \cite{espinozaandaluz2017modeling-99a,ASINARI2007359}, convolutional neural networks \cite{kishimoto2025deep-91f} or random walking method \cite{Iwai.2010, Brus.2020} directly on segmented 3D geometries.

As can be seen in the previous paragraph, the literature does not currently reflect a single, comprehensive algorithm that can predict all microstructural parameters of SOC electrodes. Each parameter is generally derived through separate techniques, equiring significant computational load, indicating an existing knowledge gap. To address this gap, we propose an all-in-one single method that enables the prediction of eight microstructural parameters of microstructure. Although demonstrated here for an Ni/YSZ anode of SOC, this methodology has significant potential for broader application, including other multiphase materials across energy, catalytic, and filtration systems.
\section{Methodology}
\subsection{Persistent homology}

Standard, voxel-based microstructure representations are often poorly suited to conventional machine learning models. A standard approach might treat the entire 3D microstructure as a raw input tensor - e.g., of shape $200 \times 200 \times 200$ - and apply convolutional neural networks (CNNs) directly. However, such an approach is both computationally inefficient and does not fully utilize the CNN capabilities due to the low-dimensional, discrete nature of phase representation. First, voxel-wise inputs are not rotation invariant, while many key microstructural features are. Second, the voxel data in SOCs are inherently categorical, typically consisting of only three discrete values that correspond to the nickel, YSZ, and pore phases. This sparse and discontinuous encoding presents a challenge for CNNs, which are less naturally suited to data where the most important structure lives in long-range, sparse, or sharply delineated patterns rather than in smoothly varying local textures. These limitations collectively impair the model's ability to generalize and extract functionally relevant features, representing a major bottleneck in applying standard machine learning techniques to microstructure analysis. To address this, we propose leveraging topological data analysis (TDA) to extract rotation and translation invariant representations, which compactly encode microstructures in a form more suitable to machine learning. Topological data analysis offers a promising and elegant approach to overcoming the limitations of voxel-based microstructure representations. Instead of relying on raw spatial encoding, TDA characterizes the \textit{shape} of data -- capturing how features like pores, particles, or channels connect and evolve across scales. Persistent homology, the backbone of TDA, was first introduced by Edelsbrunner et al. \cite{Edelsbrunner.2000} and formalized by Zomorodian and Carlsson \cite{Zomorodian.2004}. Since then, it has been widely applied in domains ranging from biomedicine to materials science \cite{Skaf.2022, Pawlowski.2023}. In the field of solid oxide cells, the persistent homology was applied to compare microstructures \cite{Yamatoko2025} or to link it with electrochemical performance \cite{Szemer.2025}, but not for feature extraction. 

At the core of persistent homology is the concept of \textit{filtration}, a process that tracks the evolution of topological properties -- such as connected components, loops, or voids -- as a filtration parameter varies. This process yields a \textit{persistence diagram}, a multiset of points where each point represents a topological feature of the original structure and encodes its \textit{birth} and \textit{death} moments \cite{Berry.2020}. These diagrams offer a compact, yet expressive summary of the multi-scale structure of 3D microstructures.
To make persistence diagrams compatible with machine learning pipelines, several vectorization methods have been developed.
One widely used technique is the \textit{persistence image} (PI), introduced by Adams et al. \cite{Adams.2016}.
This approach first converts the diagram into a smooth \textit{persistence surface}, which is then discretized into a fixed-size, continuous valued vector -- a format well suited to standard machine learning models.
PIs retain critical topological information while offering rotation and translation invariance.In particular, recent studies have demonstrated the effectiveness of persistence images in capturing key microstructural traits for SOCs and similar porous materials \cite{Obayashi.2022, Szemer.2025,Yamatoko2025}.

\begin{figure}[hb!]
    \centering
    \includegraphics{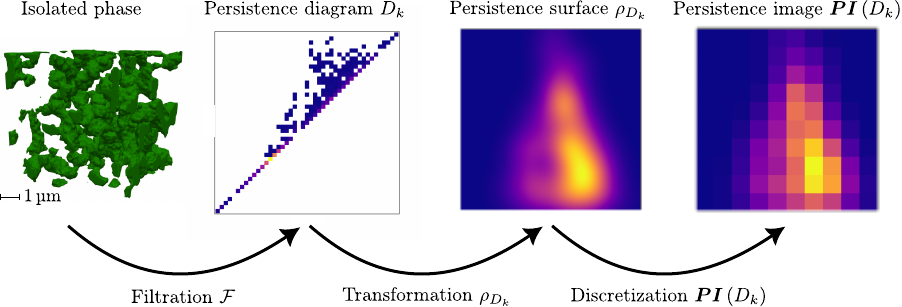}
    \caption{\textbf{Obtaining persistence image pipeline.} A selected phase is isolated, filtered to extract topological data in the form of a persistence diagram, transformed into a persistence surface, and discretized into a persistence image. Reproduced from \cite{Szemer.2025} under Creative Common licence.}
    \label{fig:topo_pipeline}
\end{figure}

Formally, persistent homology tracks the appearance and disappearance of topological features across a family of nested subsets derived from the data \cite{Berry.2020}.
A persistence surface $\rho_{D_k}(x, y)$ of a persistence diagram $D_k$ is a transformation that assigns a function over $(x, y)\in \mathbb{R}^2$ based on the persistence diagram pairs.
The transformation is given as a weighted sum of functions:
\begin{equation}
    \rho_{D_k}(x,y) = \sum_{(b_i,p_i)\in D_k}w\left(b_i,p_i\right)\phi\left(x,y,b_i,p_i\right),
\end{equation}%
where $w$ is a weight function, $\phi$ is a 
similarity kernel function, in most cases Gaussian \cite{Adams.2016, Obayashi.2022}, $b_i$ and $p_i$ are respectively the birth and persistence of the feature $i$.
We use the weight function $w$ proposed by Kusano \cite{Kusano.2017}:
\begin{equation}
\label{eq:weight_function_sec}
    w\left(b,p\right) = \arctan\left(C p^\gamma\right),
\end{equation}
with $C > 0$ and $\gamma \in \mathbb{R}_{>0}$ being parameters, which we treat as hyperparameters in this study.
Finally, the persistence image (PI) is obtained by integrating a persistence surface on elements of a grid.
For a grid element $(i, j)$ corresponding to a domain space $P_{i, j} = (a_i, a_{i+1}] \times (a_j, a_{j+1}]$, with $a_i$ defining $i$-th grid line position, the persistence image value is given as:
\begin{equation}
\label{eq:persistence_image_sec}
    PI(D_k)_{i,j}=\iint_{P_{i,j}}\rho_{D_k}\left(x,y\right)\mathrm{d}x\mathrm{d}y.
\end{equation}
The whole process of transforming raw microstructure into a persistence image is schematically illustrated on \autoref{fig:topo_pipeline}.

\subsection{Dataset}
The dataset consists of 1312 artificially generated microstructures. Each microstructure consists of 200x200x200 voxels which cover a microstructure region in the form of a cube with a side of \SI{7.14}{\micro\m}. In order to prepare the microstructures, we have used a cellular automata-based algorithm that simulates the mixing of powder and sintering process. The microstructures are originally generated in the cubic \SI{10}{\micro\m} domain and cropped to ensure microstructural homogeneity, as the generation method incorporates variations near the boundary of the domain. Details about the microstructure generation algorithm can be found in our previous work \cite{Buchaniec.2019, Szemer.2025, Prokop2024wim}.

\begin{figure}
    \centering
    \includegraphics{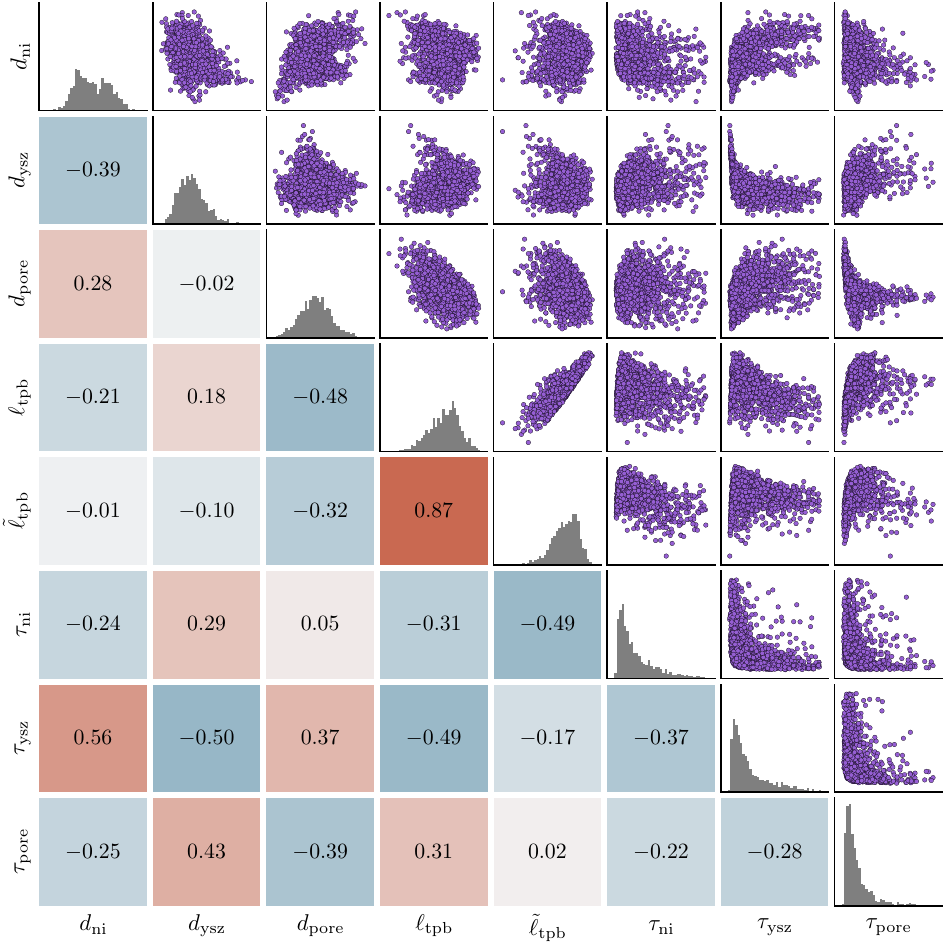}
      \caption{\textbf{Pairwise exploration of microstructural descriptors.} 
      Diagonal: histograms of each descriptor. 
      Above the diagonal: scatterplots for each pair of descriptors. 
      Below the diagonal: Pearson correlation coefficients $r$. 
      Variables: $d_{\mathrm{ni}}$—Ni grain diameter; $d_{\mathrm{ysz}}$—YSZ grain diameter; $d_{\mathrm{pore}}$—pore diameter; 
      $\ell_{\mathrm{tpb}}$—total TPB length density; $\tilde\ell_{\mathrm{tpb}}$—active TPB length density; 
      $\tau_{\mathrm{ni}}$, $\tau_{\mathrm{ysz}}$, $\tau_{\mathrm{pore}}$—tortuosity of Ni, YSZ, and pore phases, respectively}
    \label{fig:grid_plot}
\end{figure}

Each microstructure is quantitatively analyzed using several algorithms. The tortuosity factor of each phase is estimated with the use of open-source TauFactor software \cite{Cooper2016}. Individual grains are labeled using the watershed algorithm. The average grain size is calculated by averaging equivalent diameters of each grain in the given phase. Each voxel is set to be percolated or disconnected based on the possibility of reaching the cubical domain boundary by a path through neighboring voxels (von Neumann neighborhood) of the same phase. The triple phase boundary is extracted as a connected line that consists of edges which are part of voxels of each of the three phases simultaneously. The points of the surface and the TPB line that are part of the microstructure's boundary are set to be fixed. Two Laplacian smoothing iterations with relaxation factor set to $0.4$ are performed on the surface, and an additional one iteration of Laplacian smoothing is performed on the TPB line with relaxation factor $0.5$. The triple phase boundary length is approximated by the total length of all line segments after smoothing. Active triple phase boundary length includes only the segments that were part of only connected voxels.

We summarized both the univariate distributions and the pairwise relationships between the key microstructural descriptors in \autoref{fig:grid_plot}.
The diagonal panels show histograms that reveal differing scales and shapes across descriptors, with several exhibiting skewness and long tails, supporting the need for appropriate transformations prior to model training.
The upper-triangular panels provide scatterplots, while the lower-triangular panels report Pearson correlation coefficients.
The correlation panels reveal several significant relationships among microstructural features that support the choice of predicting the active TPB length density as a pretext task in the optimization process.
Specifically, there is a strong positive correlation between the active $\tilde{\ell}_{\mathrm{tpb}}$  and total $\ell_{\mathrm{tpb}}$ ($r=0.87$), indicating that models trained to predict one are likely to develop transferable representations useful for the other.
Furthermore, $\tilde\ell_{\mathrm{tpb}}$ exhibits moderate to strong correlations with other key descriptors such as the pore tortuosity ($\tau_{\mathrm{ni}}$, $r=-0.49$), and the pore size ($d_{\mathrm{pore}}$, $r=-0.32$), suggesting its central role in capturing essential microstructural patterns.
These correlations collectively reinforce the hypothesis that fine-tuning the hyperparameters based on TPB prediction allows the model to acquire a deeper understanding of the microstructure, enhancing the prediction of the other microstructural properties in subsequent tasks.

To ensure consistency across all microstructural attributes for subsequent comparisons, we applied the same preprocessing pipeline to each attribute.
First, to reduce skewness in the data, we applied the Yeo-Johnson \cite{Yeo.2000} transformation followed by $z$-score standardization.
Next, we rescaled the transformed features to the $[0, 1]$ range.
These preprocessing steps were not only beneficial for the model performance but also facilitated a consistent and meaningful comparison of how different models performed across various attributes.

We adopted a per-attribute modeling strategy, meaning that each microstructural attribute was modeled individually rather than jointly with others.
This approach helps prevent inter-attribute interference, where learning one property could negatively affect the accuracy of another.
By focusing on each attribute separately, we isolate the learning process, which reduces interference and leads to improved overall performance and more precise predictions.

\subsection{Model architecture details}
In our previous study, we introduced a method for grouping persistence images (PIs) by Betti number order to predict SOFC electrode characteristics \cite{Szemer.2025}.
The current work builds upon and formalizes this approach into a unified architecture named \texttt{NINDEN} (\textbf{N}ine \textbf{In}put \textbf{De}nse \textbf{N}etwork), incorporating architectural enhancements and training refinements to improve predictive accuracy and model generalization for higher fidelity attribute prediction.
\texttt{NINDEN} retains the core design from our earlier architecture, in which nine distinct input branches correspond to combinations of topological degree (Betti number order $k \in \{0, 1, 2\}$) and microstructural phases (e.g., nickel, YSZ, pore).
Each branch independently processes a set of persistence images related to a specific phase and Betti number, allowing the model to learn topological features of $k$-th order in a structurally aware manner.

\begin{figure}
    \centering
    \includegraphics{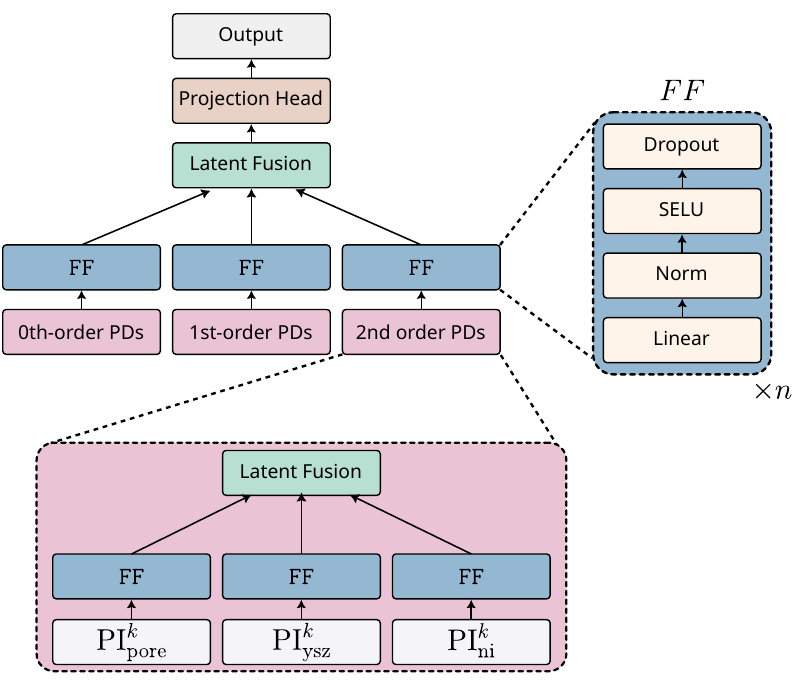}
    \caption{\textbf{Overview of the NINDEN architecture.} The model consists of multiple parallel branches, each processing persistence images through deep feed-forward (\texttt{FF}) blocks.
    Each \texttt{FF}-block includes a linear layer, followed by normalization, activation, and dropout.
    Latent vectors from all branches are fused, via concatenation in this work, before being passed to the projection head, which comprises additional fully connected layres.
    The architecture incorporates batch normalization throughout and is designed to enhance expressiveness and generalization for SOFC electrode attribute prediction
    }
    \label{fig:model_architecture}
\end{figure}

Several key modifications differentiate \texttt{NINDEN} from the original design.
First, each branch has been deepened, allowing for more expressive feature extraction.
Second, batch normalization is introduced after each dense layer to stabilize the learning rate and accelerate convergence.
Third, instead of a single dense layer, the projection head is expanded into a stack of three dense layers, each with normalization, enhancing the model's ability to capture non-linear, attribute-specific interactions in the latent space.
A schematic overview of the updated model architecture is presented on \autoref{fig:model_architecture}.

Each feed-forward block, denoted as \texttt{FF} in the architecture schema, consists of a linear layer followed by a normalization layer, an activation function, and dropout for regularization.
The latent vectors produced by these \texttt{FF}-blocks are subsequently aggregated through a latent fusion mechanism.
In this work, we employ a simple concatenation strategy for fusion, but we suggest that more advanced approaches -- such as normalization and blending techniques, or the incorporation of gating mechanisms, could be explored.
These would allow the branches to act as specialized experts, likely enabling the use of a Mixture-of-Experts (MoE) framework for potentially improved performance and interpretability \cite{Mu.2025}.
Such modularization of network architecture allows individual branches to extract potentially meaningful information during the training process, without being overshadowed by the most dominant persistence images.

\subsection{Hyperparamter optimization}

Traditional hyperparameter optimization methods, such as Grid Search and Random Search, are simple to implement but often inefficient and limited in scope \cite{Bergstra.2012}.
In contrast, Bayesian Optimization (BO) is a principled strategy for optimizing black-box functions, offering improved performance while mitigating reliance on stochastic exploration \cite{Bergstra.2011}.
By leveraging a surrogate model and an acquisition function, BO efficiently guides the search for optimal solutions.
It is both theoretically grounded and widely adopted in deep learning, particularly in scenarios where traditional optimization methods are inefficient or computationally prohibitive \cite{Bergstra.2011, Ozaki.2020}.
In our case, computational cost was a significant concern, as the generation of persistence images during each iteration of hyperparameter optimization proved highly resource intensive.
In this study, we use one of the state-of-the-art hyperparameter optimization algorithms, Tree-Structured Parzen Estimation (TPE), which models the probability density of good and bad hyperparameter configurations separately to guide the search more effectively \cite{Bergstra.2011}.

Due to the high cost of generating persistence images, to identify an optimal configuration of hyperparameters, we conducted the HPO with a strategy structured in two phases:  
\begin{enumerate}
    \item Persistence Image Optimization and Hyperspace Constraint: We used Random Search for \num{50} trials to optimize PI parameters and to constrain the space of hyperparameters of the neural network. This choice was made to mitigate the slow start problem of TPE, which requires a broad exploration before it can make meaningful predictions.
    In order to reduce the optimization complexity, parameter $\gamma$ was selected withing positive integer domain.
    
    \item Focused Hyperparameter Optimization Using TPE: With optimized persistence image parameters and a constrained model search space, we employed TPE for \num{50} trials to fine-tune the most significant hyperparameters, as identified by feature importance analysis using fANOVA\cite{Hutter.2014}.
\end{enumerate}

As a pretext task during the optimization process, we selected the prediction of active TPB ($\tilde\ell_{tpb}$).
This task encourages the model to extract deeper semantic representations from the persistence images.

The search space includes the PI parameters, the learning rate, the number of layers, and the neurons within them.
The best configuration yielded a validation MSE of \num{1.05e-02}. The details of the hyperspace, scale, and optimal values are presented in \autoref{app:hpo}.

\subsection{Training details}
The dataset was divided into training (\SI{65}{\percent}), validation (\SI{15}{\percent}), and test (\SI{20}{\percent}) subsets.
We used the SELU \cite{Klambauer.2017} activation function throughout the network and optimized the model parameters using the AdamW \cite{Loshchilov.2019} optimizer with a weight decay of \num{1.72e-1}.
To prevent overfitting, dropout layers with a rate of \num{1.2e-3} were inserted after each dense block.
Early stopping was applied based on the validation loss with patience of $4$ epochs.
Furthermore, we used a cosine annealing learning rate schedule with warm restarts \cite{Loshchilov.2017}, starting with an initial learning rate of \num{1.2e-3} and an initial restart cycle length of $25$ epochs, which was multiplied by factor $2$ after each restart.

\section{Results}

\subsection{Representative model}
To evaluate the performance and robustness of the proposed architecture, we've trained separate \texttt{NINDEN} models for each microstructural attribute under the final selected hyperparameter configuration.
Each model was trained across 10 independent runs, each with a different random initialization and data shuffling.
This procedure provides a reliable estimation of the performance variability and ensures that the reported metrics are representative of the model's generalization capability.
The average performance and standard deviation across these runs are summarized in \autoref{tab:representative_model}.
Model performance was assessed using five complementary metrics: mean squared error (MSE), mean absolute error (MAE), coefficient of determination ($R^2$), Pearson correlation coefficient ($r$), and Spearman correlation coefficient ($\rho$).
These metrics collectively offer a balanced view of both prediction accuracy and the quality of monotonic relationships between predicted and true values.

\begin{table}
    \centering
    \caption{\textbf{Predictive performance of models on the test set for each microstructural attribute.}
Each row corresponds to a model trained independently to predict a specific target attribute.
Metrics reported are the mean across $10$ independent training runs using different random seeds.
Evaluation metrics include mean squared error (MSE), mean absolute error (MAE), coefficient of determination ($R^2$), Pearson correlation coefficient ($r$), and Spearman rank correlation coefficient ($\rho$)
}
    \label{tab:representative_model}

\begin{tabular}{llllll}
\toprule
Model & MSE & MAE & $R^2$ & $r$ & $\rho$ \\
\midrule
\texttt{NINDEN }$d_{\rm{ni}}$ & $ \num{5.08e-03}$ & $ \num{5.54e-02}$ & $ \num{0.85}$ & $ \num{0.93}$ & $ \num{0.94}$ \\
\texttt{NINDEN }$d_{\rm{pore}}$ & $ \num{5.49e-03}$ & $ \num{5.90e-02}$ & $ \num{0.84}$ & $ \num{0.92}$ & $ \num{0.91}$ \\
\texttt{NINDEN }$d_{\rm{ysz}}$ & $ \num{8.07e-03}$ & $ \num{7.13e-02}$ & $ \num{0.73}$ & $ \num{0.86}$ & $ \num{0.84}$ \\
\texttt{NINDEN }$\tau_{\mathrm{ni}}$ & $ \num{9.91e-03}$ & $ \num{7.82e-02}$ & $ \num{0.84}$ & $ \num{0.92}$ & $ \num{0.94}$ \\
\texttt{NINDEN }$\tau_{\mathrm{pore}}$ & $ \num{6.15e-03}$ & $ \num{6.04e-02}$ & $ \num{0.86}$ & $ \num{0.94}$ & $ \num{0.94}$ \\
\texttt{NINDEN }$\tau_{\mathrm{ysz}}$ & $ \num{5.94e-03}$ & $ \num{6.10e-02}$ & $ \num{0.91}$ & $ \num{0.96}$ & $ \num{0.96}$ \\
\texttt{NINDEN }$\ell_{\mathrm{tpb}}$ & $ \num{4.89e-03}$ & $ \num{5.59e-02}$ & $ \num{0.88}$ & $ \num{0.94}$ & $ \num{0.93}$ \\
\texttt{NINDEN }$\tilde{\ell}_{\mathrm{tpb}}$ & $ \num{7.57e-03}$ & $ \num{6.83e-02}$ & $ \num{0.81}$ & $ \num{0.90}$ & $ \num{0.90}$ \\
\bottomrule
\end{tabular}

\end{table}

\begin{figure}[ht]
    \centering
    \includegraphics{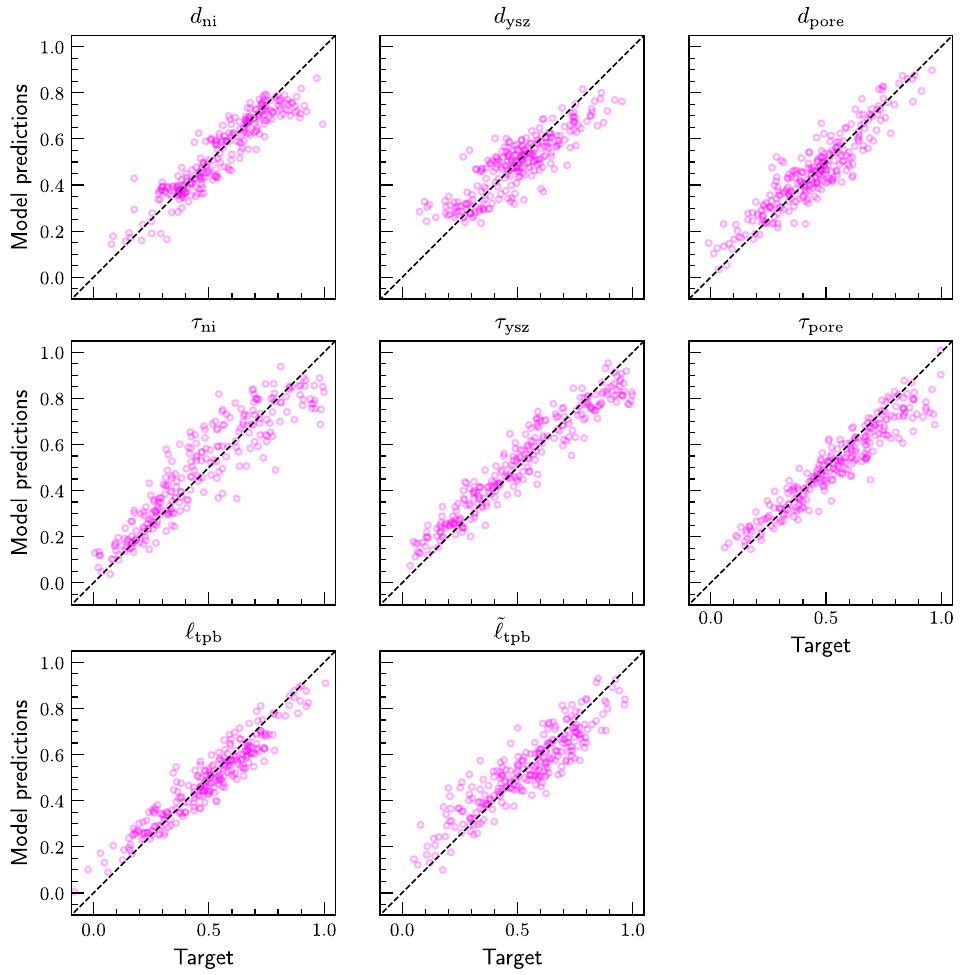}
    \caption{\textbf{Predicted vs. ground-truth values for each microstructural attribute.}
    Each scatter plot corresponds to the model whose performance was closest to the mean across $10$ runs.
    The horizontal axis shows the scaled target values, and the vertical axis shows the corresponding model predictions.
    Each pink point represents an individual observation. Diagonal represents ideal prediction
}
    \label{fig:scatters}
\end{figure}

Overall, the models achieved high predictive performance across all target attributes.
The $R^2$ values ranged from $0.73$ for $d_{\rm{ysz}}$ to $0.91$ for $\tau_{\rm{ysz}}$, indicating a strong proportion of explained variance.
High Pearson and Spearman coefficients (both exceeding $0.90$ for most targets) further confirm that the models effectively capture linear and rank-based relationships.
Notably, the model predicting active TPB length ($\tilde\ell_{\mathrm{tpb}}$) achieved $R^2 = 0.81$ with $r = 0.90$ and $\rho = 0.90$, which is competitive despite the increased complexity associated with this attribute.
Despite being trained independently, \texttt{NINDEN} models consistently generalized across various morphological patterns, demonstrating robustness and adaptability to different microstructural characteristics.
To visually assess prediction quality, \autoref{fig:scatters} present scatter plots comparing predicted values to ground-truth targets for the model from each group, whose performance was closest to the mean.
The horizontal axis represents the ground-truth target values, while the vertical axis shows the corresponding predictions made by the model, each pink point denotes an individual observation.
The observed clustering of points along the diagonal in the plots further validates the predictive fidelity of the models.

\subsection{Ablation studies}
We further conducted ablation studies by retraining models without individual phases.
To assess the effect of individual component ablations, we performed statistical significance testing using Welch's independent $t$-test, which does not assume equal variance between compared groups.
This test was applied to compare the ablated variant against the full model.
Results in \autoref{tab:phases_pruning} are annotated with \texttt{*}, \texttt{**}, and \texttt{***}, denoting significance at $10\%$, $5\%$, and $1\%$ levels, respectively.
These thresholds correspond to significant, statistically significant, and highly statistically significant differences in performance.
As shown on the \autoref{tab:phases_pruning}, excluding either the pore or YSZ phase caused only moderate degeneration in predictive performance across most metrics.
For instance, the $R^2$ values for $\ell_{\mathrm{tpb}}$ dropped from $0.88$ (full model) to $0.84$ and $0.82$ when pores and YSZ were removed, respectively, with differences being statistically significant at the \SI{5}{\percent} or \SI{1}{\percent} level.
Similarly, attributes like $d_{\mathrm{ni}}$ and $\tau_{\mathrm{pore}}$ remained highly predictable even when certain phase inputs were withheld, suggesting that the learned representations captured sufficient information from remaining phases.
However, removal of the nickel phase led to a marked deterioration in performance across several key targets.
The predictive accuracy for $\tilde\ell_{\mathrm{tpb}}$ dropped substantially ($R^2 = 0.69$, $r=0.84$), and metrics for $\tau_{\mathrm{ni}}$ and $\ell_{\mathrm{tpb}}$ also declined significantly, both statistically and practically.
This highlights the critical role of the nickel phase in determining transport-relevant features of the microstructure.
Interestingly, the model's ability to predict attributes such as $d_{\mathrm{pore}}$ and $\tau_{\mathrm{ysz}}$ remained robust even after ablation, reflecting the potential redundancy or shared encoding of phase-interaction information.
Such behavior underscores the value of encoding spatial interactions between phases, which is naturally captured by our framework's design and the underlying use of topological data analysis.
A detailed breakdown of performance differences, along with corresponding $p$-values, Cohen's $d$ effect size, and \SI{95}{\percent} confidence intervals for all comparisons is provided in the \autoref{app:ablation}.

\begin{table}
    \centering
    \caption{\textbf{Performance impact of removing individual phase branches from the model.}
    Each metric reports mean over $10$ independent training runs.
    Statistical significance versus the full model was assessed using Welch's t-test. \texttt{*}, \texttt{**}, and \texttt{***} indicate significance at the $10\%$, $5\%$, and $1\%$ levels respectively
    }
    \label{tab:phases_pruning}

\begin{tabular}{lllllll}
\toprule
Model &   & MSE & MAE & $R^2$ & $r$ & $\rho$ \\
\midrule
\multirow[t]{8}{*}{W/O pores} 
 & $d_{\mathrm{ni}}$ & $\num{5.19e-03}$ & $\num{5.58e-02}$ & $\num{0.85}$ & $\num{0.92}$ & {\boldmath$\num{0.93}$}\sig{0.05} \\
 & $d_{\mathrm{pore}}$ & $\num{5.93e-03}$ & $\num{6.01e-02}$ & $\num{0.83}$ & {\boldmath$\num{0.91}$}\sig{0.01} & {\boldmath$\num{0.89}$}\sig{0.01} \\
 & $d_{\mathrm{ysz}}$ & $\num{7.65e-03}$ & $\num{6.91e-02}$ & $\num{0.74}$ & $\num{0.87}$ & $\num{0.84}$ \\
 & $\tau_{\mathrm{ni}}$ & {\boldmath$\num{1.19e-02}$}\sig{0.05} & {\boldmath$\num{8.57e-02}$}\sig{0.05} & {\boldmath$\num{0.81}$}\sig{0.05} & {\boldmath$\num{0.91}$}\sig{0.1} & $\num{0.92}$ \\
 & $\tau_{\mathrm{pore}}$ & {\boldmath$\num{8.70e-03}$}\sig{0.01} & {\boldmath$\num{7.07e-02}$}\sig{0.01} & {\boldmath$\num{0.81}$}\sig{0.01} & {\boldmath$\num{0.91}$}\sig{0.01} & {\boldmath$\num{0.92}$}\sig{0.01} \\
 & $\tau_{\mathrm{ysz}}$ & $\num{6.08e-03}$ & $\num{6.19e-02}$ & $\num{0.91}$ & $\num{0.96}$ & $\num{0.96}$ \\
 & $\ell_{\mathrm{tpb}}$ & {\boldmath$\num{6.26e-03}$}\sig{0.05} & {\boldmath$\num{6.25e-02}$}\sig{0.1} & {\boldmath$\num{0.84}$}\sig{0.05} & {\boldmath$\num{0.92}$}\sig{0.05} & $\num{0.92}$ \\
 & $\tilde{\ell}_{\mathrm{tpb}}$ & {\boldmath$\num{9.51e-03}$}\sig{0.01} & {\boldmath$\num{7.74e-02}$}\sig{0.01} & {\boldmath$\num{0.76}$}\sig{0.01} & {\boldmath$\num{0.88}$}\sig{0.01} & {\boldmath$\num{0.88}$}\sig{0.01} \\
\cline{1-7}
\multirow[t]{8}{*}{W/O ysz}
 & $d_{\mathrm{ni}}$ & {\boldmath$\num{5.86e-03}$}\sig{0.1} & $\num{5.87e-02}$ & {\boldmath$\num{0.83}$}\sig{0.1} & {\boldmath$\num{0.91}$}\sig{0.01} & {\boldmath$\num{0.93}$}\sig{0.01} \\
 & $d_{\mathrm{pore}}$ & $\num{6.64e-03}$ & $\num{6.45e-02}$ & $\num{0.81}$ & $\num{0.91}$ & $\num{0.89}$ \\
 & $d_{\mathrm{ysz}}$ & {\boldmath$\num{9.43e-03}$}\sig{0.01} & {\boldmath$\num{7.70e-02}$}\sig{0.05} & {\boldmath$\num{0.68}$}\sig{0.01} & {\boldmath$\num{0.83}$}\sig{0.01} & {\boldmath$\num{0.80}$}\sig{0.01} \\
 & $\tau_{\mathrm{ni}}$ & {\boldmath$\num{1.25e-02}$}\sig{0.05} & {\boldmath$\num{8.83e-02}$}\sig{0.05} & {\boldmath$\num{0.80}$}\sig{0.05} & {\boldmath$\num{0.91}$}\sig{0.1} & $\num{0.92}$ \\
 & $\tau_{\mathrm{pore}}$ & $\num{5.55e-03}$ & $\num{5.75e-02}$ & $\num{0.88}$ & $\num{0.94}$ & $\num{0.94}$ \\
 & $\tau_{\mathrm{ysz}}$ & {\boldmath$\num{1.06e-02}$}\sig{0.01} & {\boldmath$\num{7.96e-02}$}\sig{0.01} & {\boldmath$\num{0.84}$}\sig{0.01} & {\boldmath$\num{0.92}$}\sig{0.01} & {\boldmath$\num{0.92}$}\sig{0.01} \\
 & $\ell_{\mathrm{tpb}}$ & {\boldmath$\num{7.19e-03}$}\sig{0.01} & {\boldmath$\num{6.84e-02}$}\sig{0.01} & {\boldmath$\num{0.82}$}\sig{0.01} & {\boldmath$\num{0.91}$}\sig{0.01} & {\boldmath$\num{0.90}$}\sig{0.01} \\
 & $\tilde{\ell}_{\mathrm{tpb}}$ & {\boldmath$\num{1.05e-02}$}\sig{0.01} & {\boldmath$\num{8.09e-02}$}\sig{0.01} & {\boldmath$\num{0.74}$}\sig{0.01} & {\boldmath$\num{0.86}$}\sig{0.01} & {\boldmath$\num{0.84}$}\sig{0.01} \\
\cline{1-7}
\multirow[t]{8}{*}{W/O nickel}
 & $d_{\mathrm{ni}}$ & $\num{5.53e-03}$ & $\num{5.72e-02}$ & $\num{0.84}$ & $\num{0.92}$ & $\num{0.93}$ \\
 & $d_{\mathrm{pore}}$ & $\num{5.36e-03}$ & $\num{5.80e-02}$ & $\num{0.84}$ & $\num{0.92}$ & $\num{0.91}$ \\
 & $d_{\mathrm{ysz}}$ & $\num{8.09e-03}$ & $\num{7.12e-02}$ & $\num{0.72}$ & $\num{0.86}$ & $\num{0.83}$ \\
 & $\tau_{\mathrm{ni}}$ & {\boldmath$\num{1.18e-02}$}\sig{0.01} & {\boldmath$\num{8.35e-02}$}\sig{0.05} & {\boldmath$\num{0.81}$}\sig{0.01} & {\boldmath$\num{0.91}$}\sig{0.01} & {\boldmath$\num{0.92}$}\sig{0.01} \\
 & $\tau_{\mathrm{pore}}$ & $\num{5.75e-03}$ & $\num{5.91e-02}$ & $\num{0.87}$ & $\num{0.94}$ & $\num{0.94}$ \\
 & $\tau_{\mathrm{ysz}}$ & $\num{6.13e-03}$ & $\num{6.21e-02}$ & $\num{0.91}$ & $\num{0.96}$ & $\num{0.96}$ \\
 & $\ell_{\mathrm{tpb}}$ & {\boldmath$\num{8.94e-03}$}\sig{0.01} & {\boldmath$\num{7.24e-02}$}\sig{0.01} & {\boldmath$\num{0.78}$}\sig{0.01} & {\boldmath$\num{0.89}$}\sig{0.01} & {\boldmath$\num{0.87}$}\sig{0.01} \\
  & $\tilde{\ell}_{\mathrm{tpb}}$ & {\boldmath$\num{1.22e-02}$}\sig{0.01} & {\boldmath$\num{8.50e-02}$}\sig{0.01} & {\boldmath$\num{0.69}$}\sig{0.01} & {\boldmath$\num{0.84}$}\sig{0.01} & {\boldmath$\num{0.84}$}\sig{0.01} \\
\bottomrule
\end{tabular}
\end{table}

\section{Discussion}

\texttt{NINDEN} introduces a modular, purely topology-based neural network architecture that leverages topological data analysis to predict structural properties in complex, multiphase microstructures.
Our results confirm that topological descriptors, when embedded within a task-specific deep learning architecture, provide a powerful inductive bias for learning structure-property mappings.

One of the core findings is the high predictive fidelity achieved across diverse microstructural attributes, with strong performance maintained even under stochastic training variations.
The ablation studies further validate the architecture's modular design: removing phase-specific inputs (e.g., pores, YSZ) led to graceful degradation in targets associated with phases' interfacial complexity.
This suggests that topological summaries encode not only the geometry of individual phases but also their spatial interactions, even under incomplete information.

From a methodological perspective, this work demonstrates how domain-inspired priors, such as topology and modularity, can be injected into deep learning pipelines to improve generalization in settings with structured but heterogeneous inputs.
Unlike purely end-to-end models, \texttt{NINDEN} offers interpretability at the representation level via topological summaries and modular attribution.

Compared to convolutional neural networks (CNNs), the proposed architecture estimates the tortuosity factor using two orders of magnitude fewer training samples while simultaneously handling higher-resolution volumes. For instance, a recent state of the art research utilizes a convolutional neural network (CNN) to estimate the microstructural parameters of a solid oxide fuel cell anode\cite{kishimoto2025deep-91f}. In the study, a CNN processing a $64 \times 64 \times 64$ voxel cube (100 nm per voxel) ingests 262,144 input features and was trained on $\approx$15,000 structures \cite{kishimoto2025deep-91f}. In contrast, our approach operates on $200 \times 200 \times 200$ voxel cubes at a resolution of 36 nm, capturing roughly six times more physical detail, but requires only 1,300 structures. This substantial gain in sample efficiency results from the machine learning assembly procedure, which reduces the $200 \times 200 \times 200$ voxel cubes with categorical data (3 values) to $32 \times 32$ persistent images with contiunous values range.

Overall, the modular \texttt{NINDEN} approach offers a generalizable framework for predicting structure-property relationships in other porous materials, leveraging persistent homology.
Our results collectively support the effectiveness of the proposed framework in learning compact and transferable representations of microstructure, paving the way for downstream tasks such as optimization and material design.
Furthermore, the results highlight the potential of persistent homology and modular representation learning in bridging the gap between structured scientific data and data-driven modeling, paving the way for interpretable, generalizable models in computational materials and beyond.

\section*{Author contributions}
\textbf{M.S.}: Methodology, Software, Validation, Formal analysis, Investigation, Data curation, Writing, Visualization
\textbf{S.B.}: Methodology, Software (Microstructure generation), Validation, Formal analysis, Investigation, Data curation, Writing, Visualization
\textbf{G.B.}: Conceptualization, Supervision, Writing, Project administration

\section*{Competing interests}
All authors declare no financial or nonfinancial competing interests. 

\section*{Declaration of generative AI and AI-assisted technologies in the writing process}
During the preparation of this work, the authors used LanguageTool, Grammarly, Writeful, and ChatGPT to refine English writing. After using these tools, the authors reviewed and edited the content as needed and took full responsibility for the content of the publication.

\section*{Data availability}
The dataset generated and analyzed during the current study is available in the \href{https://doi.org/10.58032/AGH/MQTCKT}{Cracow Open Research Data Repository}. 

\section*{Code availability}
All Python code for this study is available on \href{https://doi.org/10.58032/AGH/MQTCKT}{Cracow Open Research Data Repository}. 

\section*{Acknowledgements}
The authors acknowledge support by the program ``Excellence Initiative---Research University'' for the AGH University of Krakow, Poland, and the Research Subsidy of the AGH University of Krakow for the Faculty of Energy and Fuels. We gratefully acknowledge Polish high-performance computing infrastructure PLGrid (HPC Center: ACK Cyfronet AGH) for providing computer facilities and support within computational grant no. PLG/2025/018341

\bibliographystyle{ieeetr}
\bibliography{references}

\appendix

\section{Hyperparameter optimization}\label{app:hpo}

\begin{figure}
    \centering
    \includegraphics{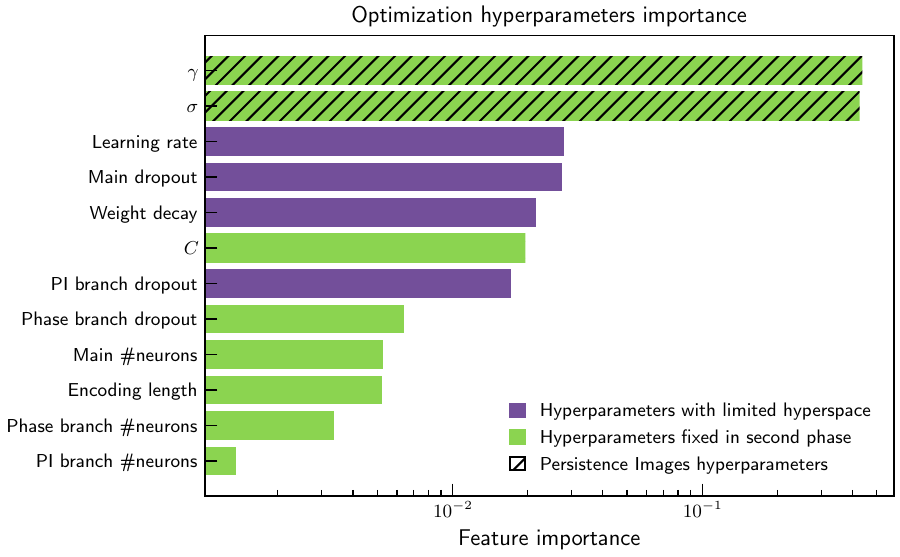}
    \caption{\textbf{Feature importances of hyperparameters estimated using fANOVA.}
    Hyperparameters from the first-stage random search are shown in orange, those selected for second-stage optimization by the Tree-Structured Parzen Estimator algorithm are highlighted in light blue, with persistence image hyperparameters additionally hatched
    }
    \label{fig:feature_importance}
\end{figure}

The feature importances are presented on \autoref{fig:feature_importance}.
Features importance were estimated using fANOVA\cite{Hutter.2014}, and are plotted on a logarithmic scale along the $x$-axis to capture the wide range of sensitivity among hyperparameters.
Hyperparameters selected for optimization in the second stage by Tree-Structured Parzen Estimator algorithm are highlighted in purple, while those corresponding to persistence images are additionally marked with hatching.
Hyperparameters defined during the first-stage random search are shown in green.
Notably, two of the PI hyperparameters, $p$ which govern the weighing function in the persistence surface, and $\sigma$ which control its smoothness, emerged as the most influential.
Their importances are approximately an order of magnitude greater than those of the remaining hyperparameters, indicating their critical role in shaping the model's representational capacity and downstream performance.
Details of the hyperparameter search space, including sampling strategies and value ranges, are provided in \autoref{tab:app_hpo_phase_1} and \autoref{tab:app_hpo_phase_2} for phase I and phase II of optimization, respectively.

\begin{table}[!ht]
    \caption{\textbf{HPO search space during first phase.}
    This initial stage employed a broad random search across architectural and training-related parameters, including the structure of both the main and auxiliary branches, dropout probabilities, persistence image encoding parameters, and optimizer settings.
    Sampling strategies include uniform and logarithmic distributions, with step sizes specified where applicable.
    The goal of this phase was to coarsely explore a wide design space and identify promising regions for more refined tuning in phase II.
    }
    \label{tab:app_hpo_phase_1}
    \centering

\begin{tabular}{llcc}
\toprule
Parameter & Sampling & Hyperparameter Value \\
\midrule
$C$ & uniform & $[ \num{0.5}, \num{40} ]$ \\
$\sigma$ & logarithmic &$[ \num{1e-3}, \num{5e-2} ]$ \\
$p$ & uniform &$[ \num{1}, \num{6} ]$ \\
Encoding length & logarithmic & $[ \num{4}, \num{7} ]$ \\
Learning rate & uniform & $[ \num{1e-3}, \num{1e-2} ]$ \\
Main \#neurons & logarithmic & $[ \num{5}, \num{10} ]$ \\
Main dropout prob & uniform  & $[ \num{0.1}, \num{0.4} ]$ \\
PI branch \#neurons & logarithmic & $[ \num{7}, \num{9} ]$ \\
PI branch dropout prob & uniform  & $[ \num{0.1}, \num{0.4} ]$ \\
Phase branch \#neurons & logarithmic  & $[ \num{7}, \num{10} ]$ \\
Phase branch dropout prob & uniform   & $[ \num{0.1}, \num{0.4} ]$ \\
Weight decay & uniform &  $[ \num{1e-2}, \num{1} ]$ \\
\bottomrule
\end{tabular}

\end{table}

\begin{table}[!ht]
    \centering
    \caption{\textbf{HPO search space during second phase.}
This stage focused on fine-tuning a reduced set of parameters within the most promising ranges identified in Phase I.
The Tree-structured Parzen Estimator (TPE) algorithm guided the search process.
Notably, categorical choices (e.g., activation functions) were included alongside continuous-valued parameters, enabling architectural flexibility while narrowing the optimization scope.
    }
    \label{tab:app_hpo_phase_2}

\begin{tabular}{llc}
\toprule
Parameter & Sampling & Hyperparamter Value \\
\midrule
Activation & logarithmic & $[$\texttt{SELU}, \texttt{GELU}, \texttt{Mish}$]$ \\
Encoding length & uniform & $[ \num{3}, \num{6} ]$ \\
Learning rate & uniform & $[ \num{1e-3}, \num{1e-2} ]$ \\
Main dropout prob & uniform & $[ \num{0.2}, \num{0.4} ]$ \\
PI branch dropout prob & uniform & $[ \num{0.2}, \num{0.4} ]$ \\
$T_{mult}$ & uniform & $[ \num{1}, \num{5} ]$ \\
Weight decay & uniform & $[ \num{0.1}, \num{1.0} ]$ \\
\bottomrule
\end{tabular}

\end{table}

\section{Details of Ablation Studies}\label{app:ablation}

\begin{table}
\caption{\textbf{Statistical comparison of ablation models with the full model using $R^2$.}
Each row reports the $p$-value from Welch’s $t$-test, significance level, mean performance difference ($\Delta\mu$), and Cohen’s $d$ effect size for a specific attribute}
\label{tab:r2}
\centering

\begin{tabular}{lllrr}
\toprule
Model & Attribute &  $p$ value  &  $\Delta\mu$  &  $d$ \\
\midrule
\multirow[t]{8}{*}{W/O nickel} 
 & $d_{\mathrm{ni}}$ & $\num{4.58e-01}$ & $\num{-1.31e-02}$ & $\num{3.41e-01}$ \\
 & $d_{\mathrm{pore}}$ & $\num{7.93e-01}$ & $\num{3.75e-03}$ & $\num{-1.19e-01}$ \\
 & $d_{\mathrm{ysz}}$ & $\num{9.62e-01}$ & $\num{-8.55e-04}$ & $\num{2.18e-02}$ \\
 & $\tau_{\mathrm{ni}}$ & $\num{9.15e-04}$ & $\num{-3.05e-02}$ & $\num{1.77e+00}   \quad \quad \quad \;$ \\
 & $\tau_{\mathrm{pore}}$ & $\num{4.13e-01}$ & $\num{8.93e-03}$ & $\num{-3.77e-01}$ \\
 & $\tau_{\mathrm{ysz}}$ & $\num{7.87e-01}$ & $\num{-2.77e-03}$ & $\num{1.23e-01}$ \\
 & $\ell_{\mathrm{tpb}}$ & $\num{1.40e-03}$ & $\num{-1.02e-01}$ & $\num{1.75e+00}   \quad \quad \quad \;$ \\
 & $\tilde{\ell}_{\mathrm{tpb}}$ & $\num{2.82e-06}$ & $\num{-1.18e-01}$ & $\num{3.41e+00}   \quad \quad \quad \;$ \\
\cline{1-5}
\multirow[t]{8}{*}{W/O pores} 
 & $d_{\mathrm{ni}}$ & $\num{7.56e-01}$ & $\num{-3.29e-03}$ & $\num{1.42e-01}$ \\
 & $d_{\mathrm{pore}}$ & $\num{2.95e-01}$ & $\num{-1.28e-02}$ & $\num{4.87e-01}$ \\
 & $d_{\mathrm{ysz}}$ & $\num{2.81e-01}$ & $\num{1.42e-02}$ & $\num{-4.97e-01}$ \\
 & $\tau_{\mathrm{ni}}$ & $\num{4.47e-02}$ & $\num{-3.09e-02}$ & $\num{1.00e+00}   \quad \quad \quad \;$ \\
 & $\tau_{\mathrm{pore}}$ & $\num{5.18e-04}$ & $\num{-5.63e-02}$ & $\num{1.89e+00}   \quad \quad \quad \;$ \\
 & $\tau_{\mathrm{ysz}}$ & $\num{8.60e-01}$ & $\num{-2.12e-03}$ & $\num{7.98e-02}$ \\
 & $\ell_{\mathrm{tpb}}$ & $\num{4.34e-02}$ & $\num{-3.46e-02}$ & $\num{9.91e-01}$ \\
 & $\tilde{\ell}_{\mathrm{tpb}}$ & $\num{6.66e-04}$ & $\num{-4.89e-02}$ & $\num{1.86e+00}   \quad \quad \quad \;$ \\
\cline{1-5}
\multirow[t]{8}{*}{W/O ysz} 
 & $d_{\mathrm{ni}}$ & $\num{6.15e-02}$ & $\num{-2.26e-02}$ & $\num{9.03e-01}$ \\
 & $d_{\mathrm{pore}}$ & $\num{2.36e-01}$ & $\num{-3.37e-02}$ & $\num{5.58e-01}$ \\
 & $d_{\mathrm{ysz}}$ & $\num{4.72e-03}$ & $\num{-4.62e-02}$ & $\num{1.45e+00}   \quad \quad \quad \;$ \\
 & $\tau_{\mathrm{ni}}$ & $\num{4.57e-02}$ & $\num{-4.05e-02}$ & $\num{1.01e+00}   \quad \quad \quad \;$ \\
 & $\tau_{\mathrm{pore}}$ & $\num{2.93e-01}$ & $\num{1.33e-02}$ & $\num{-4.85e-01}$ \\
 & $\tau_{\mathrm{ysz}}$ & $\num{3.27e-05}$ & $\num{-6.91e-02}$ & $\num{2.46e+00}   \quad \quad \quad \;$ \\
 & $\ell_{\mathrm{tpb}}$ & $\num{2.86e-03}$ & $\num{-5.77e-02}$ & $\num{1.58e+00}   \quad \quad \quad \;$ \\
 & $\tilde{\ell}_{\mathrm{tpb}}$ & $\num{3.19e-04}$ & $\num{-7.40e-02}$ & $\num{2.14e+00}   \quad \quad \quad \;$ \\
\bottomrule
\end{tabular}

\end{table}

\begin{table}
\caption{\textbf{Statistical comparison of ablation models with the full model using MAE.}
Each row reports the $p$-value from Welch’s $t$-test, significance level, mean performance difference ($\Delta\mu$), and Cohen’s $d$ effect size for a specific attribute}
\label{tab:mae}
\centering

\begin{tabular}{lllrr}
\toprule
Model & Attribute &  $p$ value  &  $\Delta\mu$  &  $d$ \\
\midrule
\multirow[t]{8}{*}{W/O nickel} 
 & $d_{\mathrm{ni}}$ & $\num{5.84e-01}$ & $\num{1.89e-03}$ & $\num{-2.50e-01}$ \\
 & $d_{\mathrm{pore}}$ & $\num{7.25e-01}$ & $\num{-9.72e-04}$ & $\num{1.60e-01}$ \\
 & $d_{\mathrm{ysz}}$ & $\num{9.67e-01}$ & $\num{-1.01e-04}$ & $\num{1.91e-02}$ \\
 & $\tau_{\mathrm{ni}}$ & $\num{1.98e-02}$ & $\num{5.33e-03}$ & $\num{-1.14e+00}   \quad \quad \quad \;$ \\
 & $\tau_{\mathrm{pore}}$ & $\num{6.05e-01}$ & $\num{-1.38e-03}$ & $\num{2.36e-01}$ \\
 & $\tau_{\mathrm{ysz}}$ & $\num{7.67e-01}$ & $\num{1.10e-03}$ & $\num{-1.36e-01}$ \\
 & $\ell_{\mathrm{tpb}}$ & $\num{2.32e-03}$ & $\num{1.65e-02}$ & $\num{-1.59e+00}   \quad \quad \quad \;$ \\
 & $\tilde{\ell}_{\mathrm{tpb}}$ & $\num{5.07e-06}$ & $\num{1.67e-02}$ & $\num{-3.02e+00}   \quad \quad \quad \;$ \\
\cline{1-5}
\multirow[t]{8}{*}{W/O pores} 
 & $d_{\mathrm{ni}}$ & $\num{8.52e-01}$ & $\num{4.22e-04}$ & $\num{-8.55e-02}$ \\
 & $d_{\mathrm{pore}}$ & $\num{6.06e-01}$ & $\num{1.13e-03}$ & $\num{-2.36e-01}$ \\
 & $d_{\mathrm{ysz}}$ & $\num{2.53e-01}$ & $\num{-2.22e-03}$ & $\num{5.28e-01}$ \\
 & $\tau_{\mathrm{ni}}$ & $\num{3.08e-02}$ & $\num{7.59e-03}$ & $\num{-1.08e+00}   \quad \quad \quad \;$ \\
 & $\tau_{\mathrm{pore}}$ & $\num{2.51e-03}$ & $\num{1.03e-02}$ & $\num{-1.58e+00}   \quad \quad \quad \;$ \\
 & $\tau_{\mathrm{ysz}}$ & $\num{8.33e-01}$ & $\num{8.84e-04}$ & $\num{-9.57e-02}$ \\
 & $\ell_{\mathrm{tpb}}$ & $\num{8.03e-02}$ & $\num{6.64e-03}$ & $\num{-8.47e-01}$ \\
 & $\tilde{\ell}_{\mathrm{tpb}}$ & $\num{3.33e-04}$ & $\num{9.05e-03}$ & $\num{-1.98e+00}   \quad \quad \quad \;$ \\
\cline{1-5}
\multirow[t]{8}{*}{W/O ysz} 
 & $d_{\mathrm{ni}}$ & $\num{1.62e-01}$ & $\num{3.36e-03}$ & $\num{-6.63e-01}$ \\
 & $d_{\mathrm{pore}}$ & $\num{2.61e-01}$ & $\num{5.58e-03}$ & $\num{-5.27e-01}$ \\
 & $d_{\mathrm{ysz}}$ & $\num{1.00e-02}$ & $\num{5.67e-03}$ & $\num{-1.29e+00}   \quad \quad \quad \;$ \\
 & $\tau_{\mathrm{ni}}$ & $\num{4.23e-02}$ & $\num{1.01e-02}$ & $\num{-1.02e+00}   \quad \quad \quad \;$ \\
 & $\tau_{\mathrm{pore}}$ & $\num{3.65e-01}$ & $\num{-2.97e-03}$ & $\num{4.15e-01}$ \\
 & $\tau_{\mathrm{ysz}}$ & $\num{2.73e-04}$ & $\num{1.86e-02}$ & $\num{-2.08e+00}   \quad \quad \quad \;$ \\
 & $\ell_{\mathrm{tpb}}$ & $\num{3.97e-03}$ & $\num{1.25e-02}$ & $\num{-1.53e+00}   \quad \quad \quad \;$ \\
 & $\tilde{\ell}_{\mathrm{tpb}}$ & $\num{1.46e-04}$ & $\num{1.25e-02}$ & $\num{-2.23e+00}   \quad \quad \quad \;$ \\
\bottomrule
\end{tabular}

\end{table}

\begin{table}
\caption{\textbf{Statistical comparison of ablation models with the full model using MSE.}
Each row reports the $p$-value from Welch’s $t$-test, significance level, mean performance difference ($\Delta\mu$), and Cohen’s $d$ effect size for a specific attribute}
\label{tab:mse}
\centering

\begin{tabular}{lllrr}
\toprule
Model & Attribute &  $p$ value  &  $\Delta\mu$  &  $d$ \\
\midrule
\multirow[t]{8}{*}{W/O nickel} 
 & $d_{\mathrm{ni}}$ & $\num{4.58e-01}$ & $\num{4.52e-04}$ & $\num{-3.41e-01}$ \\
 & $d_{\mathrm{pore}}$ & $\num{7.93e-01}$ & $\num{-1.28e-04}$ & $\num{1.19e-01}$ \\
 & $d_{\mathrm{ysz}}$ & $\num{9.62e-01}$ & $\num{2.51e-05}$ & $\num{-2.18e-02}$ \\
 & $\tau_{\mathrm{ni}}$ & $\num{9.15e-04}$ & $\num{1.94e-03}$ & $\num{-1.77e+00}   \quad \quad \quad \;$ \\
 & $\tau_{\mathrm{pore}}$ & $\num{4.13e-01}$ & $\num{-4.04e-04}$ & $\num{3.77e-01}$ \\
 & $\tau_{\mathrm{ysz}}$ & $\num{7.87e-01}$ & $\num{1.85e-04}$ & $\num{-1.23e-01}$ \\
 & $\ell_{\mathrm{tpb}}$ & $\num{1.40e-03}$ & $\num{4.05e-03}$ & $\num{-1.75e+00}   \quad \quad \quad \;$ \\
 & $\tilde{\ell}_{\mathrm{tpb}}$ & $\num{2.82e-06}$ & $\num{4.67e-03}$ & $\num{-3.41e+00}   \quad \quad \quad \;$ \\
\cline{1-5}
\multirow[t]{8}{*}{W/O pores}
 & $d_{\mathrm{ni}}$ & $\num{7.56e-01}$ & $\num{1.14e-04}$ & $\num{-1.42e-01}$ \\
 & $d_{\mathrm{pore}}$ & $\num{2.95e-01}$ & $\num{4.36e-04}$ & $\num{-4.87e-01}$ \\
 & $d_{\mathrm{ysz}}$ & $\num{2.81e-01}$ & $\num{-4.18e-04}$ & $\num{4.97e-01}$ \\
 & $\tau_{\mathrm{ni}}$ & $\num{4.47e-02}$ & $\num{1.96e-03}$ & $\num{-1.00e+00}   \quad \quad \quad \;$ \\
 & $\tau_{\mathrm{pore}}$ & $\num{5.18e-04}$ & $\num{2.54e-03}$ & $\num{-1.89e+00}   \quad \quad \quad \;$ \\
 & $\tau_{\mathrm{ysz}}$ & $\num{8.60e-01}$ & $\num{1.41e-04}$ & $\num{-7.98e-02}$ \\
 & $\ell_{\mathrm{tpb}}$ & $\num{4.34e-02}$ & $\num{1.38e-03}$ & $\num{-9.91e-01}$ \\
  & $\tilde{\ell}_{\mathrm{tpb}}$ & $\num{6.66e-04}$ & $\num{1.94e-03}$ & $\num{-1.86e+00}   \quad \quad \quad \;$ \\
\cline{1-5}
\multirow[t]{8}{*}{W/O ysz}
 & $d_{\mathrm{ni}}$ & $\num{6.15e-02}$ & $\num{7.80e-04}$ & $\num{-9.03e-01}$ \\
 & $d_{\mathrm{pore}}$ & $\num{2.36e-01}$ & $\num{1.15e-03}$ & $\num{-5.58e-01}$ \\
 & $d_{\mathrm{ysz}}$ & $\num{4.72e-03}$ & $\num{1.36e-03}$ & $\num{-1.45e+00}   \quad \quad \quad \;$ \\
 & $\tau_{\mathrm{ni}}$ & $\num{4.57e-02}$ & $\num{2.58e-03}$ & $\num{-1.01e+00}   \quad \quad \quad \;$ \\
 & $\tau_{\mathrm{pore}}$ & $\num{2.93e-01}$ & $\num{-6.03e-04}$ & $\num{4.85e-01}$ \\
 & $\tau_{\mathrm{ysz}}$ & $\num{3.27e-05}$ & $\num{4.61e-03}$ & $\num{-2.46e+00}   \quad \quad \quad \;$ \\
 & $\ell_{\mathrm{tpb}}$ & $\num{2.86e-03}$ & $\num{2.30e-03}$ & $\num{-1.58e+00}   \quad \quad \quad \;$ \\
  & $\tilde{\ell}_{\mathrm{tpb}}$ & $\num{3.19e-04}$ & $\num{2.94e-03}$ & $\num{-2.14e+00}   \quad \quad \quad \;$ \\
\bottomrule
\end{tabular}

\end{table}

\begin{table}
\caption{\textbf{Statistical comparison of ablation models with the full model using Pearson correlation coefficient.}
Each row reports the $p$-value from Welch’s $t$-test, significance level, mean performance difference ($\Delta\mu$), and Cohen’s $d$ effect size for a specific attribute}
\label{tab:p_corr}
\centering

\begin{tabular}{lllrr}
\toprule
Model & Attribute &  $p$ value  &  $\Delta\mu$  &  $d$ \\
\midrule
\multirow[t]{8}{*}{W/O nickel} 
 & $d_{\mathrm{ni}}$ & $\num{1.90e-01}$ & $\num{-8.57e-03}$ & $\num{6.12e-01}$ \\
 & $d_{\mathrm{pore}}$ & $\num{6.93e-01}$ & $\num{-2.16e-03}$ & $\num{1.80e-01}$ \\
 & $d_{\mathrm{ysz}}$ & $\num{4.95e-01}$ & $\num{-6.98e-03}$ & $\num{3.16e-01}$ \\
 & $\tau_{\mathrm{ni}}$ & $\num{3.87e-04}$ & $\num{-1.67e-02}$ & $\num{1.95e+00}   \quad \quad \quad \;$ \\
 & $\tau_{\mathrm{pore}}$ & $\num{5.74e-01}$ & $\num{3.19e-03}$ & $\num{-2.58e-01}$ \\
 & $\tau_{\mathrm{ysz}}$ & $\num{3.59e-01}$ & $\num{-3.88e-03}$ & $\num{4.25e-01}$ \\
 & $\ell_{\mathrm{tpb}}$ & $\num{5.56e-04}$ & $\num{-5.23e-02}$ & $\num{1.92e+00}   \quad \quad \quad \;$ \\
 & $\tilde{\ell}_{\mathrm{tpb}}$ & $\num{1.45e-06}$ & $\num{-6.52e-02}$ & $\num{3.66e+00}   \quad \quad \quad \;$ \\
\cline{1-5}
\multirow[t]{8}{*}{W/O pores} 
 & $d_{\mathrm{ni}}$ & $\num{1.47e-01}$ & $\num{-6.35e-03}$ & $\num{6.82e-01}$ \\
 & $d_{\mathrm{pore}}$ & $\num{5.80e-03}$ & $\num{-1.10e-02}$ & $\num{1.40e+00}   \quad \quad \quad \;$ \\
 & $d_{\mathrm{ysz}}$ & $\num{1.65e-01}$ & $\num{6.97e-03}$ & $\num{-6.47e-01}$ \\
 & $\tau_{\mathrm{ni}}$ & $\num{6.53e-02}$ & $\num{-1.58e-02}$ & $\num{9.14e-01}$ \\
 & $\tau_{\mathrm{pore}}$ & $\num{1.04e-03}$ & $\num{-2.35e-02}$ & $\num{1.83e+00}   \quad \quad \quad \;$ \\
 & $\tau_{\mathrm{ysz}}$ & $\num{6.46e-01}$ & $\num{-2.50e-03}$ & $\num{2.09e-01}$ \\
 & $\ell_{\mathrm{tpb}}$ & $\num{4.40e-02}$ & $\num{-1.82e-02}$ & $\num{9.82e-01}$ \\
 & $\tilde{\ell}_{\mathrm{tpb}}$ & $\num{6.55e-04}$ & $\num{-2.76e-02}$ & $\num{1.90e+00}   \quad \quad \quad \;$ \\
\cline{1-5}
\multirow[t]{8}{*}{W/O ysz} 
 & $d_{\mathrm{ni}}$ & $\num{3.46e-03}$ & $\num{-1.54e-02}$ & $\num{1.51e+00}   \quad \quad \quad \;$ \\
 & $d_{\mathrm{pore}}$ & $\num{1.15e-01}$ & $\num{-1.46e-02}$ & $\num{7.68e-01}$ \\
 & $d_{\mathrm{ysz}}$ & $\num{3.52e-04}$ & $\num{-3.40e-02}$ & $\num{2.10e+00}   \quad \quad \quad \;$ \\
 & $\tau_{\mathrm{ni}}$ & $\num{7.17e-02}$ & $\num{-1.57e-02}$ & $\num{8.90e-01}$ \\
 & $\tau_{\mathrm{pore}}$ & $\num{5.22e-01}$ & $\num{4.47e-03}$ & $\num{-2.92e-01}$ \\
 & $\tau_{\mathrm{ysz}}$ & $\num{4.31e-06}$ & $\num{-3.94e-02}$ & $\num{2.96e+00}   \quad \quad \quad \;$ \\
 & $\ell_{\mathrm{tpb}}$ & $\num{3.38e-03}$ & $\num{-2.88e-02}$ & $\num{1.55e+00}   \quad \quad \quad \;$ \\
 & $\tilde{\ell}_{\mathrm{tpb}}$ & $\num{2.86e-04}$ & $\num{-4.14e-02}$ & $\num{2.20e+00}   \quad \quad \quad \;$ \\
\bottomrule
\end{tabular}

\end{table}

\begin{table}
\caption{\textbf{Statistical comparison of ablation models with the full model using Spearman correlation coefficient.}
Each row reports the $p$-value from Welch’s $t$-test, significance level, mean performance difference ($\Delta\mu$), and Cohen’s $d$ effect size for a specific attribute}
\label{tab:s_corr}
\centering

\begin{tabular}{lllrr}
\toprule
Model & Attribute &  $p$ value  &  $\Delta\mu$  &  $d$ \\
\midrule
\multirow[t]{8}{*}{W/O nickel} 
 & $d_{\mathrm{ni}}$ & $\num{2.86e-01}$ & $\num{-7.48e-03}$ & $\num{4.99e-01}$ \\
 & $d_{\mathrm{pore}}$ & $\num{9.31e-01}$ & $\num{6.37e-04}$ & $\num{-3.91e-02}$ \\
 & $d_{\mathrm{ysz}}$ & $\num{8.30e-01}$ & $\num{-2.25e-03}$ & $\num{9.84e-02}$ \\
 & $\tau_{\mathrm{ni}}$ & $\num{3.15e-03}$ & $\num{-1.33e-02}$ & $\num{1.52e+00}   \quad \quad \quad \;$ \\
 & $\tau_{\mathrm{pore}}$ & $\num{5.73e-01}$ & $\num{3.52e-03}$ & $\num{-2.58e-01}$ \\
 & $\tau_{\mathrm{ysz}}$ & $\num{4.49e-01}$ & $\num{-3.00e-03}$ & $\num{3.49e-01}$ \\
 & $\ell_{\mathrm{tpb}}$ & $\num{2.12e-04}$ & $\num{-6.37e-02}$ & $\num{2.09e+00}   \quad \quad \quad \;$ \\
 & $\tilde{\ell}_{\mathrm{tpb}}$ & $\num{2.73e-06}$ & $\num{-6.47e-02}$ & $\num{3.58e+00}   \quad \quad \quad \;$ \\
\cline{1-5}
\multirow[t]{8}{*}{W/O pores} 
 & $d_{\mathrm{ni}}$ & $\num{4.28e-02}$ & $\num{-7.59e-03}$ & $\num{9.75e-01}$ \\
 & $d_{\mathrm{pore}}$ & $\num{2.77e-03}$ & $\num{-1.70e-02}$ & $\num{1.57e+00}   \quad \quad \quad \;$ \\
 & $d_{\mathrm{ysz}}$ & $\num{1.66e-01}$ & $\num{8.19e-03}$ & $\num{-6.47e-01}$ \\
 & $\tau_{\mathrm{ni}}$ & $\num{1.22e-01}$ & $\num{-1.32e-02}$ & $\num{7.48e-01}$ \\
 & $\tau_{\mathrm{pore}}$ & $\num{4.12e-03}$ & $\num{-2.11e-02}$ & $\num{1.54e+00}   \quad \quad \quad \;$ \\
 & $\tau_{\mathrm{ysz}}$ & $\num{6.85e-01}$ & $\num{-2.11e-03}$ & $\num{1.84e-01}$ \\
 & $\ell_{\mathrm{tpb}}$ & $\num{1.99e-01}$ & $\num{-1.34e-02}$ & $\num{6.01e-01}$ \\
 & $\tilde{\ell}_{\mathrm{tpb}}$ & $\num{1.51e-03}$ & $\num{-2.40e-02}$ & $\num{1.72e+00}   \quad \quad \quad \;$ \\
\cline{1-5}
\multirow[t]{8}{*}{W/O ysz}
 & $d_{\mathrm{ni}}$ & $\num{1.32e-03}$ & $\num{-1.39e-02}$ & $\num{1.70e+00}   \quad \quad \quad \;$ \\
 & $d_{\mathrm{pore}}$ & $\num{2.92e-01}$ & $\num{-1.11e-02}$ & $\num{4.93e-01}$ \\
 & $d_{\mathrm{ysz}}$ & $\num{5.84e-04}$ & $\num{-3.84e-02}$ & $\num{2.04e+00}   \quad \quad \quad \;$ \\
 & $\tau_{\mathrm{ni}}$ & $\num{1.22e-01}$ & $\num{-1.29e-02}$ & $\num{7.46e-01}$ \\
 & $\tau_{\mathrm{pore}}$ & $\num{5.14e-01}$ & $\num{4.77e-03}$ & $\num{-2.98e-01}$ \\
 & $\tau_{\mathrm{ysz}}$ & $\num{3.70e-06}$ & $\num{-3.88e-02}$ & $\num{3.04e+00}   \quad \quad \quad \;$ \\
 & $\ell_{\mathrm{tpb}}$ & $\num{4.46e-03}$ & $\num{-3.36e-02}$ & $\num{1.50e+00}   \quad \quad \quad \;$ \\
 & $\tilde{\ell}_{\mathrm{tpb}}$ & $\num{1.53e-04}$ & $\num{-5.60e-02}$ & $\num{2.49e+00}   \quad \quad \quad \;$ \\
\bottomrule
\end{tabular}

\end{table}

\end{document}